**A Si-memristor electronically and uniformly switched by a constant voltage**

Yang Lu and I-Wei Chen

Department of Materials Science and Engineering, University of Pennsylvania, Philadelphia, Electronic mail: iweichen@seas.upenn.edu

Amorphous insulators have localized wave functions that decay with the distance $r$ following $\exp(-r/\zeta)$. Since nanoscale conduction is not excluded at $r<\zeta$, one may use amorphous insulators and take advantage of their size effect for nanoelectronic applications. Voltage-regulated nanoscale conductivity is already utilized in metal-insulator-metal devices known as memristors. But typically their tunable conductivity does not come from electrons but from migrating ions within a stoichastically formed filament, and as such their combined resistor-memory performance suffers. Here we demonstrate amorphous-silicon-based memristors can have coherent electron wave functions extending to the full device thickness, exceeding 15 nm. Remarkably, despite the large aspect ratio and very thin thickness of the device, its electrons still follow an isotropic, three-dimensional pathway, thus providing uniform conductivity at the nanometer scale. Such pathways in amorphous insulators are derived from overlapping gap states and regulated by trapped charge, which is stabilized by electron-lattice interaction; this makes the memristor exhibit pressure-triggered insulator→metal transitions. Fast, uniform, durable, low-power and purely electronic memristors with none of the shortcomings of ion-migrating memristors have been fabricated from a variety of amorphous silicon compositions and can be readily integrated into silicon technology. Therefore, amorphous **silicon may provide the ideal platform for building proximal memories, transistors and beyond.**

Next-generation silicon memory can provide a natural solution for silicon technology to meet the increasing demand for data storage. Currently, a large amount of non-volatile memory utilizes flash memory made of $SiO_2$ floating gates, which are relatively slow and power consuming. Much attention has therefore been directed to resistive random access memory—RRAM—built out of memristors[1-2] (meaning memory plus resistor) that feature a much faster switching speed (in ns), smaller write/erase voltage (1 V) and higher density (sub-10-nm size)[3-6]. Memristors constructed from myriad thin-film oxides and insulators are usually initiated by a process called forming, which creates a conducting filament in the virgin film by voltage-induced dielectric breakdown[7-9]. The filament resistance is later alternated between a low resistance state (LRS) and a high resistance state (HRS) by applying a switching voltage, which triggers a sequence of events including ion migration, redox reactions and Joule heating[7-9]. Unfortunately, the stochastic nature of ionic breakdown always leaves an imprint on the device, and because of this it is generally agreed that even the best memristor suffers from intrinsic switching/performance variability and low yield[6,10]. Moreover, the additional chemical complexity of any non-silicon memristor would complicate the task of integrating it into silicon technology, e.g., forming a 3-dimentional (3D) array of high-density on-chip memory next to the processor. Yet Si-based filamentary memristors made of $SiO_2$, $Si_3N_4$ and their suboxides/nitrides cannot provide a viable solution because they all have much inferior performance in switching voltage (higher, ~5 V), speed (slower, ~100 ns) and endurance (poorer, ~$10^4$ cycles)[11-15]; in fact, some of them only operate in reducing atmospheres[15].

An alternative to filamentary memristors is nanometallic memristors[16-19], which are all electronic and uniformly switching, and they utilize two novel features of amorphous insulators (**Fig. 1a**): (a) nanoscale electron conductivity or "nanometallicity," which manifests when the transport distance $r$ falls below the localization length $\zeta$ of the amorphous insulator[16,20], and (b) charge storage/release at easily deformable "soft spots", which are ubiquitous in amorphous structures and can regulate nanoscale conductivity by imposing/lifting long-range Coulomb blockade[18-19]. With the aid of electron doping, nanometallic memristive behavior has been engineered in many wide-band-gap amorphous oxides and nitrides[16-18]. What we will demonstrate below is that nanometallic memristors can also be realized in amorphous silicon that has a much smaller band gap (~2 eV). This finding is built on our knowledge that undoped amorphous Si is already nanometallic—its films up to 25 nm thick is conducting—though not switching[21]. But the breakthrough came from the discovery that O and N doping can open the band gap to reinforce electron trapping/detrapping at the soft spots, thus gating device current without using any filament. The resulting nanometallic Si-memristor can be easily integrated into silicon technology with unprecedentedly outstanding properties. Below we describe their properties and device physics.

**A non-filamentary, purely electronic Si-memristor**

Si-memristors of two configurations were fabricated to compare their memristive behavior. The first is a set of crossbar memristors (**Fig. 1a,** left panel) ranging from $2 \times 2$ μm$^2$ to $64 \times 64$ μm$^2$ in cell size. The memristor cell rests on a Si or quartz substrate coated by a 3 nm Ti adhesion layer followed by a patterned 25 nm Pt bottom electrode. An aligned pattern of a 2 – 7.5 nm amorphous film of O-doped Si was next deposited by reactive RF sputtering, then immediately covered—without breaking vacuum—by a 30 nm Mo top electrode array by DC sputtering. The second is a set of cells 50 – 250 µm in radius (**Fig. 1a**, right panel) made by a shadow mask method with a continuous 30 nm Mo bottom electrode and a patterned 40 nm Pt top electrode. (See **Methods** for more details.) Using the latter construct, we also fabricated other memristors using various anion doping schemes and amorphous film compositions. These include (a) $O_2$ and $N_2$ reactive sputtering or co-sputtering an oxide/nitride target (from $Si_3N_4$, AlN, $SiO_2$, $Al_2O_3$ to $HfO_2$) to introduce O and N, and (b) various semiconductor targets (from pure Si, P-Si, B-Si, Si-Ge to Ge) to alter the base composition. The final O/N concentration was determined by x-ray photoemission spectroscopy (XPS) using Si:$2p$, O:$1s$ and N:$1s$ peaks calibrated by $SiO_2$ and $Si_3N_4$ standards. Since these memristors share almost identical resistance switching characteristics, below we will mostly use those made of O-doped Si, prepared by either ($O_2$) reactive sputtering or co-sputtering Si/$Al_2O_3$ targets, in both configurations, to demonstrate their universal behavior. Data on other memristors are shown in **Supplementary figures** and elsewhere[22].

Our memristors are unambiguously non-filamentary as demonstrated by the following tests. (1) The as-fabricated crossbar devices all started from the LRS (circles in **Fig. 1b**), ready to switch to the HRS under a positive nominal voltage of ~2.5 V. (By definition, current flows from Pt to Mo under a positive voltage. Since Pt has a higher work function than Mo, electrons under a positive voltage are injected from Mo into the film.) In contrast, the virgin state of a typical filamentary memristor is highly resistive and requires forming before it can be switched. (2) The LRS has a uniform conductivity as revealed by the fracture test (inset of **Fig. 1c**), which mechanically severed a memristor cell with a resistance $R_0$ and area $A_0$ (250 × 250 µm²), preset in either LRS or HRS, into two pieces with their respective $R$ and $A$. As shown in **Fig. 1c**, for both states, $RA = R_0A_0$ is obeyed by all the severed pieces, meaning uniform conductivity at least on the scale of the smallest $A$, or several μm in length. In contrast, the same fracture test had previously spotted localized filaments in standard $HfO_2$ and $TiO_2$ filamentary memristors preset in the LRS[23-24], as it found only one severed half inheriting the resistance of the intact piece while the other half orders of magnitude more resistive. (3) A filamentary memory having a single filament perpendicular to the film must have anisotropic conductance regardless of conducting mechanism. This also holds for spurious film conductivity attributable to pinholes. However, in a magnetic field, Si-memristor has identical magnetoresistance (MR) in the parallel ($\theta=0°$) and the perpendicular ($\theta=90°$) direction relative to the field (**Fig. 1d**); indeed, the MR remains unchanged when the sample is rotated to all other directions (see inset). The mechanism and a quantitative analysis of isotropic MR are described in **Methods** along with other aspects of the resistance dependence on temperature and memory states. Nevertheless, regardless of mechanism there is no doubt that the LRS conductivity is uniform and three-dimensional. For a memristor cell with a very large lateral length (2–250 μm) but a very thin thickness (~5 nm), this requires 3D uniformity to be maintained along the latter, shorter dimension. Therefore, the conductivity must be uniform and 3D at the nanometer length scale.

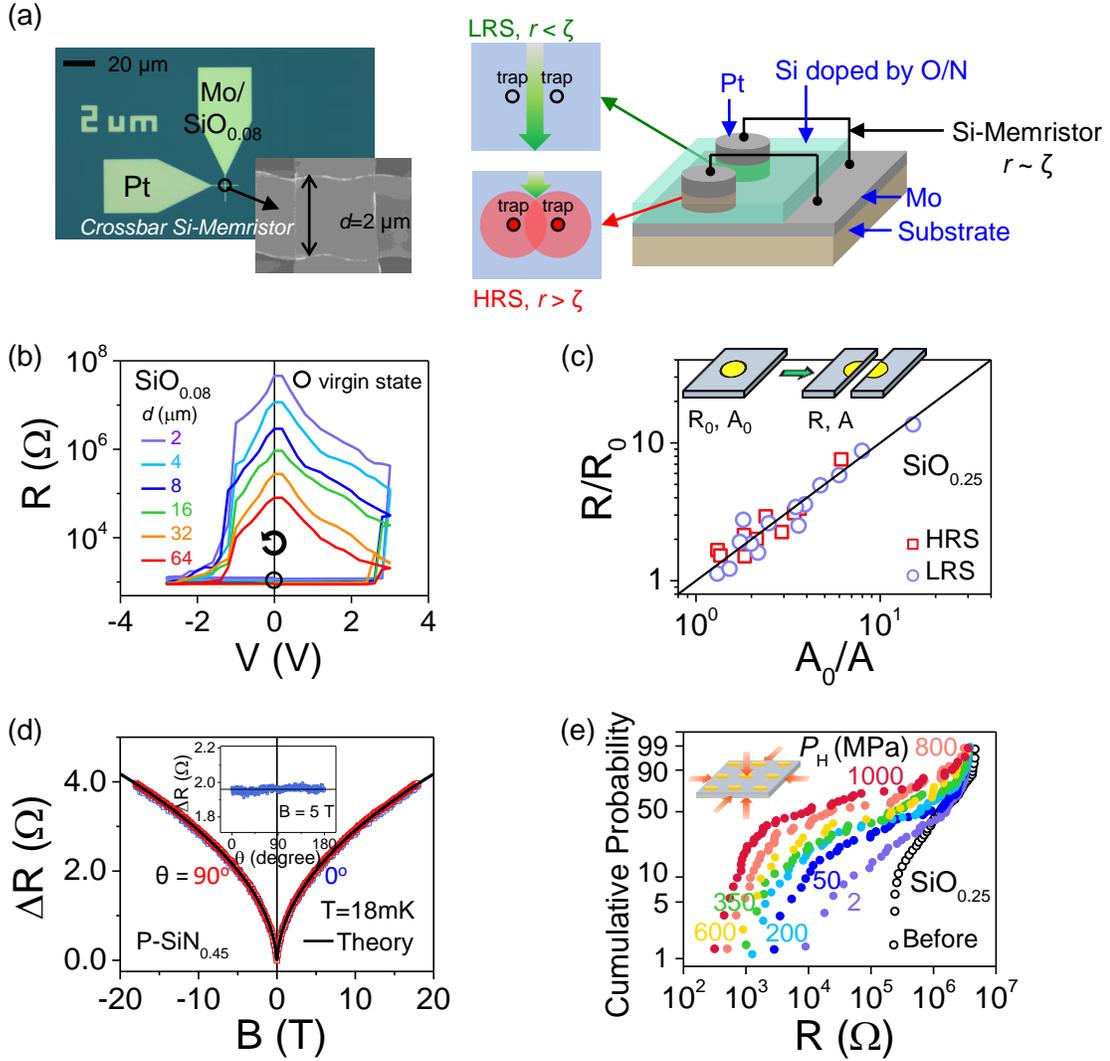

**Figure 1 Si-memristor: a non-filamentary distinctively electronic two-terminal memory-plus-resistor nano device.** (a) Left: light microscopy image of crossbar Pt/SiO$_{0.08}$/Mo memristor of $2 \times 2$ μm$^2$ (inset: enlarged SEM image). Right: similar Si-memristors fabricated by shadow mask. Center: schematic of trap-electron-regulated, electron-localization/delocalization transition from LRS (empty traps and $\zeta > r$) to HRS (filled traps and $\zeta < r$), where $\zeta$=localization length and $r \sim \delta$=film thickness. (b) Resistance ($R$) – voltage ($V$) hysteresis of SiO$_{0.08}$ crossbars (thickness: 5 nm) in various sizes tested without current compliance starting with (circled) virgin-state resistance in LRS. Arrow indicates switching direction. (c) Area dependence of resistance of fractured Mo/SiO$_{0.25}$/Pt memristors; $RA$ after fracture is the same as $R_0 A_0$ before fracture. Inset: schematic of fracture test. (d) Overlapping magnetoresistance of Mo/P-SiN$_{0.45}$/Pt memristor in LRS at 18mK in parallel and perpendicular orientations with respect to the field. Curve: fitting from theory described in **Methods**. Inset: relative changes of 5 T magnetoresistance of the same device in other orientations. (e) Cumulative Weibull probability of resistance of 50 Mo/SiO$_{0.25}$/Pt memristors pressure-switched from the preset HRS (curve labeled "Before") to lower resistance states under various hydraulic pressure $P_H$ (inset).

Unlike filamentary memristors, Si-memristors switch purely electronically. This is most evident from the pressure-triggered HRS→LRS transition, which is an insulator→metal transition. In this experiment, further described in **Methods**, the samples were disconnected from any external electrical source, and a uniform hydraulic pressure $P_H$ (**Fig. 1e,** inset) as low as a few MPa was used to trigger the transition (**Fig. 1e,** more data in **Supplementary Figure 1**.) The transition is clearly statistical because its yield, as represented by the pressure-induced resistance drop, increases with pressure and cell size. Mechanically triggered HRS→LRS transition was also achieved under a negative pressure $P_B$ generated by a magnetic impulse delivered by the passage of a single 25 × 25 × 25 μm$^3$ electron bunch, which contains 10$^9$ electrons each traveling at almost the speed of light with an energy of 22 GeV (**Supplementary Figure 2, and ref. 19**). While this impulse lasted only 10$^{-13}$ s—the time required for the relativistic electrons to travel 25 μm, it nevertheless caused all the cells within a 300 μm distance from the 25 × 25 μm$^2$ impact footprint to switch to the LRS. Importantly, the pressure-set LRS can all be voltage-switched back to the HRS, and their subsequent HRS↔LRS all proceeds at the same switching voltage and HRS/LRS resistance as if in normal voltage-switching (**Supplementary Figure 3**). Therefore, pressured caused no damage to the cells, yet it triggered a transition that must be purely electronic, since ion migration, redox reactions or Joule heating cannot possibly be driven by a pressure and proceed at such fast speed.

The non-filamentary, purely electronic Si-memristor has a constant bipolar switching voltage $V^*$, which is independent of all the device/testing variables including area, thickness, composition, temperature and switching rate. Except for the opposite polarity, $V^*$ is very similar for the HRS→LRS (SET) and LRS→HRS (RESET) transitions, even though the switching curves are asymmetric in **Fig. 1b**. This is because the device resistance $R_{LRS/HRS}$ includes both the active (Si) film resistance $R_s$ and the (passive or parasitic) load resistance $R_{load}$, which overwhelms $R_s$ in the LRS because the metallic LRS film is not at all resistive compared to the $R_{load}$ of a spreading electrode, which has a very long transport distance and a very thin thickness[25-27]. (Such $R_{load}$ also accounts for the seemingly area-insensitive LRS resistance in **Fig. 1b.**) To find the critical $V^*$ that triggers switching in the active film, we use the equivalent circuit in the inset of **Fig. 2a** to write $R_{LRS/HRS}=V^*/I_{cc}+R_{load}$, where $V^*/I_{cc}=R_s$ and $I_{cc}$ is the SET current. Plotting $R_{LRS}$ under different $I_{cc}$ from 1 μA to 2 mA in **Fig. 2a** for the 2 × 2 μm$^2$ Pt/SiO$_{0.08}$/Mo crossbar devices (**Fig. 1a**) then gives the (LRS→HRS) $V_{RESET}^*$=+1.08 V and $R_{load}$= 343 Ω. (Analysis of additional switching curves are shown in **Supplementary Figure 4.**) For the (HRS→LRS) $V_{SET}$*, the same analysis performed on $R_{HRS}$ gave a $V$* the same as read from **Fig. 1b** because, in this case, $R_{HRS} >> R_{load}$ with the Si film being insulating. These critical voltages are independent of the device area (**Fig. 2b**), the width of the switching-voltage pulse (**Fig. 2c**) and temperature (**Fig. 2d**). In **Fig. 2c**, the inset further shows that abrupt switching is triggered by a 10 ns voltage pulse upon reaching the critical voltage. Abrupt switching is also evident at 2K in **Fig. 2d**, again at the same critical voltage even though the HRS is orders of magnitude higher at 2K. Switching dynamics with such total insensitivity to time and temperature is entirely expected for a purely electronic mechanism, but not for mechanisms mediated by ion migration, redox reactions or heat generation.

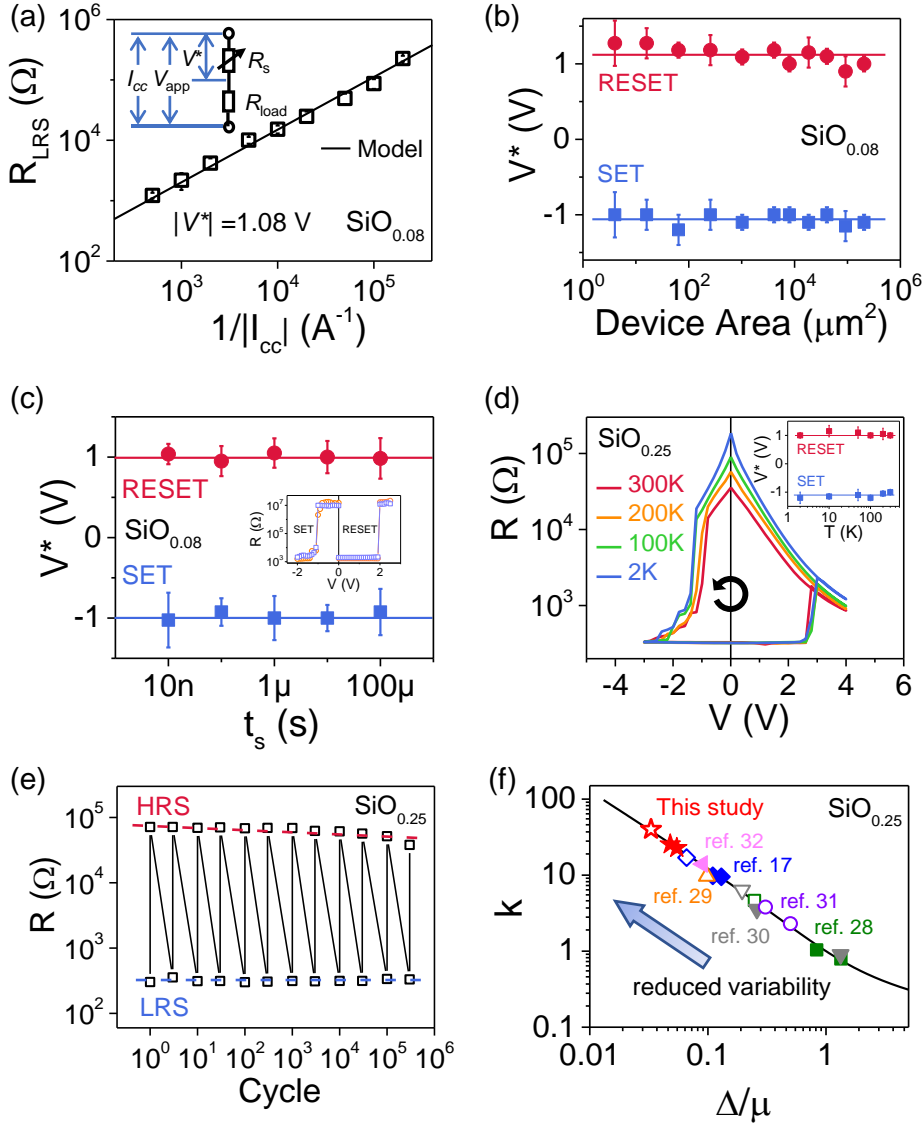

**Figure 2 Constant-voltage switching of Si-memristor.** (a) Constant critical switching voltage $V^*$ of 1 V implies linear relation between $R_{LRS}$ and $1/I_{cc}$ in $2 \times 2$ µm² Pt/SiO$_{0.08}$/Mo crossbar devices. Error bars are average of 5 tests. Inset: equivalent circuit model for $V^*$ and $R_{load}$. (b) Constant $V^*$ in RESET and SET, at ±1 V, independent of device area. Error bars are average of 10 cells of the same size. (c) Constant $V^*$ in RESET and SET, at ±1 V, for $t_s$ (voltage pulse width) from 100 µs to 10 ns for the same crossbar device. Error bars are average of 10 tests. Inset: overlapped resistance reading during pulse switching at $t_s$=100 µs (blue) and 10 ns (red). (d) $R$-$V$ switching curves of Mo/SiO$_{0.25}$/Pt memristor from 300K to 2K. Inset: constant $V^*$ vs. temperature. Error bars are average of 5 tests. (e) Endurance of HRS and LRS shows no degradation over $10^5$ cycles in $50 \times 50$ µm² × 10 nm Mo/SiO$_{0.25}$/Pt devices. (f) Weibull modulus $k$ vs standard deviation over mean $\Delta/\mu$ of switching voltage distributions (hollow symbols) and HRS/LRS resistance distributions (filled symbols), for our memristors in (e) (red stars, out of 35 cells) and state-of-the-art memistors in literature (other symbols in different colors, out of ~10-20 cells, from Ref. 17, 28-32). As pointed by arrow, a higher $k$ corresponds to a tighter distribution and smaller variations. Showing the same trend, solid curve is the known relation, $\Delta/\mu=[\Gamma(1+2/k)-\Gamma^2(1+1/k)]^{1/2}/\Gamma(1+1/k)$, between $k$ and $\Delta/\mu$ of the Weibull distribution. ($\Gamma$ is gamma function.)

In our experiments, pulse switching time was limited to 10 ns (**Fig. 2c** inset) by the parasitic RC delay. But the energy required for switching was already defined by the data, which is 10 ns × 1 V × 1 µA = 10 fJ for a 2 × 2 µm² cell. The uniform nature of switching and conduction as well as the constant $V$* suggests a scalable switching power, $P \sim IV = V^{*2}/R \propto A$. This in turn gives a scalable switching energy, which is 2.5 fJ $A(\mu m^2)$ if we express $A$ in µm² and assume 10 ns switching time. Therefore, writing a 100 × 100 nm² cell in 10 ns would require an energy 0.025 fJ or a power 2.5 nW per bit, although shorter switching time and less energy/power are possible given the fact that electron bunch can switch the cells at 0.1 ps. The robust LRS/HRS states as well as their intermediate states obtained by DC and fast-pulse switching show little cycle-to-cycle variation (**Fig. 2e**) and cell-to-cell variation (**Fig. 2f**, **Supplementary Figure 5.**) Indeed, compared to the state-of-the-art filamentary memristors with optimized uniformity (data from the literature[17,28-32]), the switching parameters of our memristors have a smaller standard-deviation-to-mean (Δ/µ) and a higher Weibull slope ($k$) as shown in **Fig.**

**2f**. Lastly, the resistance states can all be maintained over time when held at different temperatures (**Supplementary Figure 6**). Therefore, a nanometallic Si-memristor despite its purely electronic nature is unencumbered by the voltage-time dilemma:[18] It switches fast, at low voltage, yet it reliably retains resistance memory.

**Engineering localization transitions in amorphous insulators**

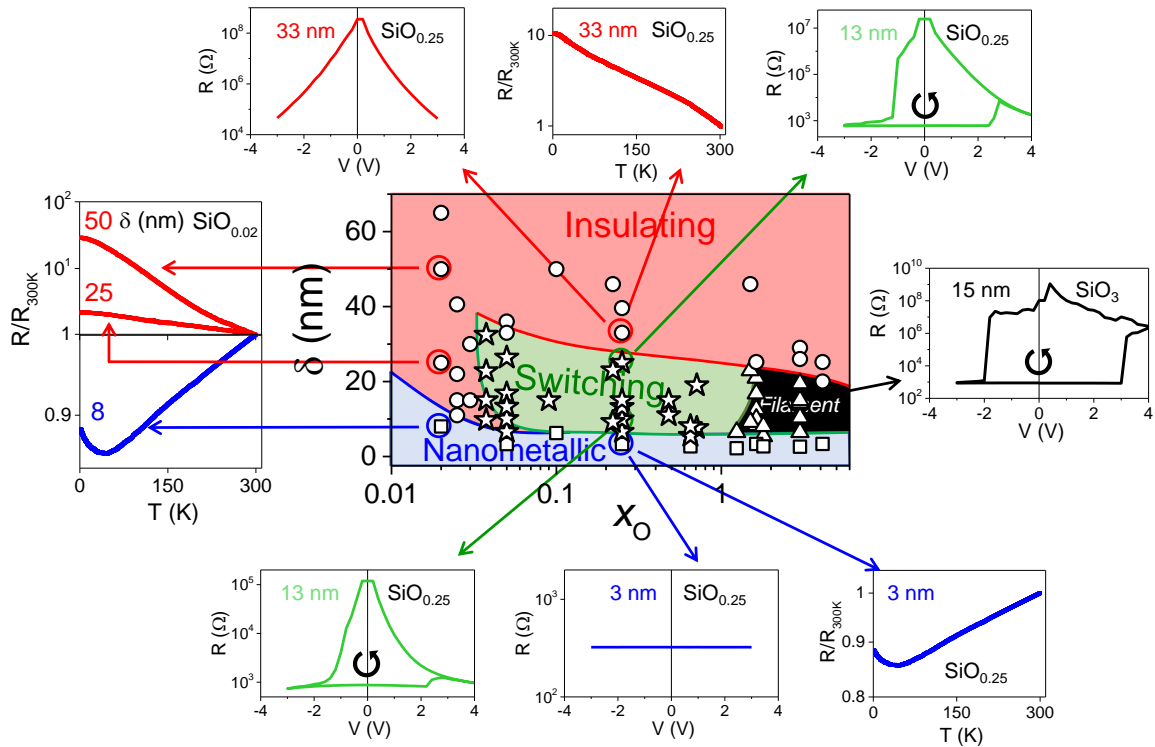

**Figure 3 Size-dependent localization transitions in amorphous SiO$_x$ nanofilms.** Center: δ - $x_O$ map of electrical properties of SiO$_x$ films made by co-sputtering Si and Al$_2$O$_3$ targets. Four types of behavior are denoted by squares for thin conductors, circles for thick insulators, stars for intermediate-thickness non-volatile nanometallic memristors starting with LRS, and triangles for electro-formed filamentary memristors starting with very high resistance. Representative *R-V* curves for each type and *R-T* curves (*R* normalized by 300K resistance) for non-switchable types are pointed to by arrows. Top electrode: Pt, bottom electrode: Mo. Devices area: 100 × 100 μm$^2$ as defined by shadow mask.

Previously, electron doping was used to make nitrides (Si$_3$N$_4$ and AlN) and oxides (SiO$_2$, Al$_2$O$_3$, MgO, Y$_2$O$_3$, HfO$_2$ and Ta$_2$O$_5$) nanometallic and switchable. For these wide band-gap insulators, the only effective method to provide itinerant electrons without resorting to phase percolation is to lightly dope them with dispersed metal atoms, Pt in the case of oxides[16,17], and Pt as well as other main group metals (e.g., Al) and transition metals (e.g., Cr) in the case of nitrides[18]. With an intermediate band gap of ~2 eV, undoped amorphous Si is an insulator with divergent resistivity at low temperatures. But at the nanoscale we found it already nanometallic,[21] behaving like a good metal when the transport length (i.e., thickness δ) is less than 10 nm (see $SiO_{0.02}$ pointed out by the blue arrow in the left of **Fig. 3**). Therefore, the reason why it is not switchable unlike the film of $SiO_{0.04}$ in the same figure is not the lack of nanometallicity but the need for more potent electron localization. This led us to O and N doping, which is known to increase electronegativity in amorphous Si thus lowering the valence band without affecting the conduction band[33]. It also led us to other similar schemes such as $Al_2O_3$ and $Si_3N_4$ doping, since Al and Si are interchangeable in the random network and O and N should have a similar doping effect. As shown in the center map of **Fig. 3**, for each intermediate film thickness δ there is indeed a range of O composition $x$ that renders the film switchable—with the virgin state in the LRS; this is the light-green region. Within this region, δ and $x$ have no effect on $V^*$, suggesting the origin of $V^*$ is energetic and not electrical field, which is consistent with the fact that these Si-memristors regardless of thickness do not require forming. The O/N-doping approach also works for amorphous films of B/N-Si and Ge—the latter has an even smaller band gap estimated to be ~1 eV—and for other compositions of amorphous semiconductor/dopant combinations, which all yielded forming-free nanometallic memristors (**Supplementary Figures 7-8**.) For completeness and to rule out the possibility of interface mechanism for nanometallic switching, also shown in **Fig. 3** are (a) a small δ blue region that is nanometallic but not switchable, e.g., $SiO_{0.25}$ of 3 nm as pointed to by the blue arrow at the bottom, (b) a large δ red region that is insulating, e.g., $SiO_{0.25}$ of 33 nm as pointed to by red arrows, and (c) a large $x$ black region that is insulating but can be rendered filamentary switching after forming, e.g., $SiO_3$ ($Si:Al_2O_3$ to be precise) as pointed to by the black arrow. Qualitatively similar demarcation of the $(x, \delta)$ space was also seen when using $N_2$ doping and/or other sputtering targets, which always include some combinations that make for a good memristor (**Supplementary Figures 7-8**).

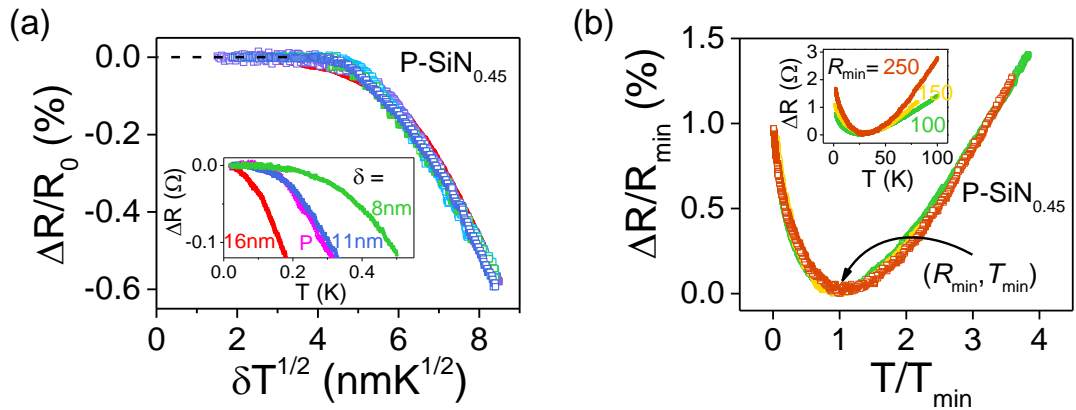

**Figure 4 Electron delocalization over sample thickness $\delta$ demonstrated by thickness-dependent saturation of quantum correction to conductivity.** (a) Normalized increase of LRS resistance of Mo/P-Si $N_{0.45}$/Pt memristor saturates at the same $\delta T^{1/2}$ for $\delta$ from 8 nm to 16 nm, with more details in inset showing two 11 nm memristors, one switched by voltage (blue), the other by pressure (purple with P), have same behavior. (b) Normalized resistance minimum ($R_{min}$) at $T_{min} \sim 35$ K for one memristor set, all from the same 11-nm-thick set in (a) but with different $I_{cc}$ to obtain different LRS $R_{min}$ (inset).

In amorphous Si, the conductance scales with the transport length $r$ (i.e., $\delta$ in our experiment) as $\exp(-r/\zeta)$, where $\zeta$ may be regarded as a localization length[21-22]. In the context of Anderson localization that envisions wave functions exponentially decay in a random insulator[34], nanometallicity is expected from the residual conductivity when $r$ falls below $\zeta$[20]. We found direct evidence of electron delocalization over the film thickness $\delta$ by studying the quantum correction to resistance (QCC) in the LRS in the same low-temperature regime where we measured prominent MR (**Fig. 1d**). In this regime, the infrequent inelastic scattering endows electrons a long coherent length $L_T$, over which their energy and phase remain invariant and their quantum interference gives rise to suppressed conductivity[35-36]. Since electrons in a random network must undergo a 3D random walk with a diffusivity $D$ and a diffusion time $\hbar/k_B T$, $L_T$ is given by the diffusion length $(\hbar D/k_B T)^{1/2}$ and is temperature-dependent[35]. It follows that below a sufficiently low temperature when $L_T > \delta$, the QCC must saturate. This is shown in **Fig. 4a** in the normalized $\delta T^{1/2}$ plot for the LRS of four Si films of different $\delta$, one switched to the LRS by pressure. These data provided unequivocal evidence that not only can LRS electrons communicate between the two electrodes along the 3D conducting paths (hence $\zeta > \delta$ in the LRS), they also interfere quantum mechanically. In Si-memristors, this QCC persists up to 35K giving rise to the resistance minimum in **Fig. 4b**.

Resistance minimum signaling the onset of QCC has been previously seen in many mesoscopic metals although their sample sizes are all too big to saturate the QCC[37]. But where are the coherent electrons from in a random insulator? Since a nanometallic memristor is made of a film of amorphous insulator sandwiched between two electrodes of different work functions ($\phi$), we propose that they come from gap-state electrons, specifically those that are energetically bracketed by $\phi_{Pt}$ and $\phi_{Mo}$ and with overlapping wave functions that extend to the full transport length $\delta$. With the Fermi level set at $\frac{1}{2}(\phi_{Pt} + \phi_{Mo})$, these states can receive from Mo tunneling electrons that populate the 3D conducting paths. However, within the same energy bracket there are also other *isolated* gap states that may receive tunneling electrons, but because they are electronically disconnected from the electrodes they must become negatively charged once tunneling electrons have arrived. With an insulating background, these charged states will impose long-range Coulomb blockade to block the device current, and if the electrons are later released, then the blockade is also lifted and the device current restored. In this way, the gap states can provide both the conducting paths and the "floating gates," so that nanometallic switching is tantamount to a delocalization/localization transition (**Fig. 1a**): The LRS corresponds to $\zeta > \delta$, the HRS to $\zeta < \delta$.

Using the above picture, we can understand why nanometallicity and switching depend on the band gap $E_g$. With a large $E_g$, the density of gap states is low, so their overlap is difficult, which makes $\zeta$ too small to reach $\delta$. Conversely, with a small $E_g$, the density of gap states is high, so few isolated gap states exist, which provide too few floating gates to regulate current. Therefore, a wide $E_g$ insulator like amorphous $SiO_2$ need electron doping, e.g., by Pt, to lower $E_g$ to render it nanometallic and switchable, and a small $E_g$ insulator like amorphous Si and Ge is already nanometallic but requires anion doping to increase $E_g$ to render it switchable. It is also plausible that the density of gap states in the insulator film is relatively similar in each optimally tuned memristor, so the switching requirements embodied by $V^*$ are relatively similar. This explains why $V^*$ of all the nanometallic memristors is always about 1 V. Obviously, the picture also foresees that nanometallic memristors cannot realize non-volatile memory if they use two identical electrodes, which is our experience in combining electrode materials.

As previously pointed out for nanometallic memristors[18-19], easily deformable soft spots ubiquitous in amorphous networks are favorable electron trapping sites because they can be locally distorted to minimize the repulsion between the extra electron and the existing electrons, which we shall call bond electrons. Strong repulsion between the bond electrons and the extra electron may be envisioned using a toy model of a molecule with a cis (or boat) configuration, which places the two sets of electrons at the two ends of the "boat." In comparison, less repulsion is felt if the boat can swivel the extra-electron end to convert itself to a "chair" configuration in which the extra electron resides at the leg, which is farther away from the bond electrons that are now at the top of the chair. Local distortions of this type at soft spots may thus stabilize trapped electrons. This toy model also explains the pressure effect: If the chair configuration stabilizes a trapped electron, then forcing it back to the boat configuration under a pressure will surely destabilize the electron and prompt its release, thus removing the Coulomb blockade. Lastly, since performing a local "molecular" conversion as if cranking an atomic-scale lever[18] requires only a time of atomic vibration, this mechanism is consistent with electron-bunch triggered sub-ps HRS→LRS switching. First-principles modeling of $ZrO_2$, $CeO_2$ and $SrTiO_3$ confirmed that a gap-state energy change of more than 1 eV can arise from a fractional local cation displacement inside an oxygen cage[19]. Therefore, a modest local atomic displacement at the soft-spots in response to the arrival of an extra electron can add much stability to soft-spot floating gates, allowing them to robustly operate in nanometallic memristors.

**Conclusions**

Finding the ~0K ground-state electrons with wave functions extended to 15 nm in an amorphous insulating film sandwiched between metal electrodes is especially significant because they can be electronically regulated, thus allowing memristive switching and possible signal amplification. A variety of amorphous Si-Ge compositions fully compatible with the silicon technology already lend themselves to such applications, forming purely electronic memristors with superior uniformity, scalability, reliability, speed and power efficiency. This portends well for the further development of Si-Ge technology to realize proximal memories and transistors for on-chip storage, in-memory computation, and possibly even biology-inspired computing.

This research used the facilities at NHMFL (DMR-1157490, State of Florida) and at FACET (SLAC National Laboratory supported by the US Department of Energy) where the experimental assistance of Drs. J-H. Park (NHMFL), H-W. Baek (NHMFL) and I. Tudosa (FACET) is gratefully acknowledged.

**Methods**

**Device fabrication:** To fabricate crossbar Si-memristors, 3 nm Ti adhesion layer and 25 nm bottom electrode of Pt were deposited consecutively by ebeam-evaporation onto a photoresist-patterned (double layered positive resists of LOR3A+S1813) silicon substrate, followed by lift-off to define the Pt bottom electrode sizes of 2-64 μm. Using the same mask, a second layer of identically patterned photoresist was next aligned—though in the perpendicular direction to the pattern of the bottom Pt layer—and deposited. Lastly, a ~5 nm layer of amorphous Si doped with O and 30 nm and a layer of Mo electrode with an additional 10 nm Pt protective overcoat were consecutively sputtered and lift-off to complete the fabrication. To fabricate vertical memristors, a (100 nm) thermal-oxide-coated 100 p-type silicon single crystal was used as the substrate, which when unheated was first coated by a 30 nm thick Mo bottom electrode using DC sputtering under a pressure of 7 mTorr. Next, after reaching a base pressure of $3 \times 10^{-7}$ Torr or less, an amorphous Si thin film with appropriate O/N dopants was RF sputter deposited without breaking the vacuum (maintained at 5 mTorr during sputtering.) Various Si targets were used in the latter process, including Si (bulk resistivity > 1 Ωcm), n-type Si (P-doped, bulk resistivity < 0.1 Ωcm), and p-type Si (B-doped, bulk resistivity ~ 0.005-0.020 Ωcm). For O/N doping during Si sputtering, the dopant was either introduced by flowing $O_2$/$N_2$ gas together with the Ar carrier gas into the sputtering chamber, or by RF co-sputtering an oxide/nitride target chosen from $Si_3N_4$, AlN, $SiO_2$, $Al_2O_3$ and $HfO_2$. Composition of O/N was tuned by the flow rate of $O_2$/$N_2$ gas or the sputtering power of the oxides/nitride target. Finally, a 40 nm thick Pt top electrode was deposited by RF sputtering (under a pressure of 7 mTorr) through a shadow mask that defined cells of 50-250 μm in radius. The spacing between the top Pt electrodes are typically 150 μm so they are electrically isolated from each other while accessing a single device by probing the top electrode (TE) of Pt and the common bottom electrode (BE) of Mo. Similar switching behaviors were also found in devices based on the Ge or $Si_{0.5}Ge_{0.5}$ composition, which were also made by RF sputtering, this time with a Ge target (Sb-doped, bulk resistivity ~ 5-40 Ωcm) or a Si-Ge alloy target.

**Sample Characterization:** To determine the film thickness, atomic force microscopy (AFM, Asylum MFP-3D) was used to measuring the step height created during deposition when one side was intentionally blocked. It was also used to measure the film roughness to check for any possible pin-holes and voids that may cause electrode contact. Such pin-holes were usually not a problem when the Si/Ge film thickness was more than 2 nm thick. X-ray photoelectron spectra (XPS) on Si films doped with O or N were collected using a RBD upgraded PHI-5000C ESCA system (Perkin Elmer) with Mg $K_\alpha$ radiation ($hv$ = 1253.6 eV). Binding energies were calibrated by referring to carbon (C 1 $s$ = 284.6 eV). Composition of O and N in Si films was then determined using Si:$2p$, O:$1s$, and N:$1s$ peaks by comparing them to calibrated signals from standard $SiO_2$ and $Si_3N_4$ samples.

**Electrical measurement and switching:** To study electrical switching of Si-memristor, current-voltage (*I-V*) and resistance-voltage (*R-V*) curves, where *R=V/I*, were measured in both continuous DC and pulsed AC modes using a Keithley 237 measurement unit and an Agilent 81104A pulse generator. In a typical setting, samples were placed in a probe station (S-1160, Signatone Corporation, Gilroy, CA). In DC measurement, a continuous sweep of the bias voltage supplied by a Keithley 237 was applied to the top Pt electrode while the bottom electrode was grounded. Current compliance during the negative sweep (on-switching from HRS to LRS or SET) was applied in order to achieve intermediate states. In AC measurement, an Agilent 81104A pulse generator was used to supply a square-impulse-shaped voltage-pulse train of a constant width but an increasing height (voltage), and a Keithley 237 was used to read the device resistance at 0.2 V after each pulse (again with bottom electrode grounded). Cell-to-cell and cycle-to-cycle variations of Si- memristor were obtained by collecting switching data under the same electrical sweeping conditions (either DC or AC) for 50 cells on the same device chip and for 100 DC cycling or over $10^5$ pulse switching cycles for a single cell. During retention tests, the resistance values of DC- and pulse-switched resistance states (including HRS, intermediate states, and LRS) were periodically recorded at room temperature after they were kept at 175 ºC for a certain time period.

**Switching by hydraulic pressure:** Before the hydraulic pressure treatment, the two-point resistance of each Si- memristor cell in the memristor array on the same chip was read at 0.1 V; if desired, they may also be preset to certain resistance state using Keithley 237. The preset states may include both the insulator state (HRS) and the metallic state (LRS), while other cells were left in their current states after reading the resistance. Next, the chip was disconnected from the voltage source, wrapped in an aluminum foil, vacuum-sealed in an elastomer bag, and suspended in a liquid-filled pressure vessel (Autoclave Engineers, Erie, US) that was charged to a preset hydraulic pressure of $P_H$ =2-350 MPa at room temperature and held for <5 min before sample removal. Some higher pressure (up to 1 GPa) experiments were also similarly performed in a hydraulic pressure vessel (Dr. CHEF) at Takasago Works, Kobe Steel, at Takasago, Japan. The resistance of each cell was read again at 0.1 V and compared with its value before the pressure treatment. The result of transition statistics in terms of post-$P_H$ resistance ranking is presented in cumulative probability curves. Lastly, pressure-induced LRS cells were tested by electrical voltage to check if they can switch back to HRS and to undergo further switching. These post-$P_H$ electrical switching curves were compared with the curves before the pressure treatment; no difference was found in the resistance and the transition voltages.

**Switching by magnetic pressure:** The magnetic pressure $P_B$, which is a negative pressure on the Si film, was generated by a burst magnetic flux that entered the Si-filled gap between the two (TE and BE) electrodes. Because of the very short duration of the burst, the flux is basically an ultrafast AC flux which generates an ultrafast AC current, so this configuration may be regarded as a metallic "container" that confines the burst magnetic field. The burst magnetic flux was received from an electron bunch, which is a spatially localized bundle of 20 GeV electrons generated at Stanford Linear Accelerator Center (SLAC) using the FACET facility. Each bunch contained ~$10^9$ electrons (>1 nC) that are narrowly collimated (~25 μm in all three directions). It had a short duration of passage, ~0.1 ps, which is the time for the bunch to travel 25 μm at near the speed of light. The bunch was available as a single bunch or as a set of repeated bunches. We usually only allowed each cell to see one bunch in our experiment; after each shot the sample was moved to a distal new position so the next shot could be seen by a set of entirely different cells. The electron bunch hit the sample chip in the normal direction. Since the maximum magnetic field around the bunch is ~ 70 T at the edge of the bunch, i.e., ~ 40 μm from the flight path, and it decays with the radial *distance* from the bunch roughly according to 1/*distance* (see **Supplementary Figure S2**), we can estimate the magnetic pressure $P_B$ in each cell from the cell location relative to the flight path using the formula $P_B[\text{bar}]=(B[\text{T}]/0.501)^2$ as shown in https://en.wikipedia.org/wiki/Magnetic_pressure. (To maximize the induced magnetic pressure inside the cell, we chose the cell size to be 20 μm, comparable to the bunch size.) The estimated maximum magnetic pressure is ~1,950 MPa, and at 500 μm away it decays to ~12.5 MPa. Before the magnetic-pressure treatment, cells were preset by a voltage to certain resistance states, with their two-point resistance values recorded at 0.1 V by a Keithley 237. Their resistance was again read in the same way after the $P_B$ treatment and compared with the pre-treatment value. To aid data presentation, each cell may be colored to indicate its resistance value before and after the treatment, and the colored maps are presented for comparison (see refs. 19 and 22).

**Low temperature switching:** Low temperature switching experiment was conducted in two systems: a Lakeshore cryogenic probe station and a Quantum Design PPMS. The former one, equipped with two probes that touch top and bottom electrodes, facilitates identical switching measurement in the S-1160 station at the temperature range of 77K to 400K (under $10^{-7}$ Torr level vacuum) by purging the sample stage with liquid nitrogen while electrically heating the stage simultaneously. In PPMS, its Ever-Cool helium cooling system allows temperate control from 2K to 300K in a vacuum of ~5 Torr. Si- memristors were mounted on a special chip holder with heat conducting vacuum grease, and gold wires were bonded (dia. = 2 μm) to the device electrodes and to the pins on the sample holder by silver paint. Below 10K, an appropriate current compliance was carefully chosen to prevent a sudden temperature burst due to Joule heating by the switching current.

**Low temperature conductivity and magnetoresistance measurement:** Temperature dependent electrical properties of memristors and similar but non-switching devices were measured with and without a magnetic field in (a) PPMS (2-300K, –9-9 T) and (b) two superconducting magnets at National High Magnetic Field Laboratory (SCM2, 0.3-2K, –18-18 T; and SCM1, 0.018-2K, –18-18 T). Devices of different Si or Ge thickness or preset in different resistance states were mounted on sample holders, either a sample puck for PPMS or a set of 16-pin dip sockets for SCM1 and SCM2. With one end soldered to the pins on the sample holder, a gold wire was silver-paste-bonded to the top Pt electrode of a Si/Ge cell of a radius of 250 μm. Another wire was similarly bonded to one edge of the bottom Mo electrode to form a two-terminal connection to the cell. Alternatively, a three-terminal connection was formed by further bonding a third gold wire to the opposite edge of the Mo electrode; these three-terminal connections used samples deposited on a fused silica substrate, which is insulating. During measurements, Three-point AC resistance was measured using an SR 830 lock-in amplifier and a standard resistor of 100 kΩ. The amplifier sent a sine wave of 1 V amplitude at 31 Hz to the standard resistor and the cell, in serial connection. Since the resistance of our cell in the metallic state is typically much less than 100 kΩ, the above is equivalent to applying a constant current of 10 μA across one lead of the top electrode and one side of the bottom electrode. Meanwhile, using a lock-in amplifier, the voltage (A-B voltage) between the other lead of the top electrode and the other side of the bottom electrode was measured when locked to the same frequency of 31 Hz at a time constant of 1 s. An auto offset and 10× signal expansion was further used to improve the voltage resolution to enable measuring small resistance changes. Electrical measurements were mainly conducted under two temperature/field-sweep conditions: cooling/heating at a fixed magnetic field (often zero-field), and ramping magnetic field at a fixed temperature. During these measurements, synchronized voltage, current, temperature and field data were recorded while the heating/cooling rate was set at an appropriate value. Sweeping of the magnetic field was typically at 0.5 T/min in PPMS and 0.3 T/min in SCM1 and SCM2. The field/sample-orientation dependence of magnetoresistance was determined in SCM1 and SCM2 by rotating the sample at 3 degrees/min in a fixed magnetic field, or by rotating the sample to a new orientation, then sweeping the field to ± 18T. Precaution against temperature/magnetic-field/orientation/voltage hysteresis was taken by repeating all the temperature/magnetic-field/orientation/voltage sweeps from a reference temperature/field/voltage both in increasing value and in decreasing value; only ramping curves that exactly overlapped in the two directions were used for analysis.

**Data analysis of quantum corrections and magnetoresistance of Si-memristor**: (For further details, see ref. 38.) In data analysis, a resistance-different method was utilized to obtain the intrinsic film response to the temperature/field perturbation without being contaminated by the parasitic signal of the electrodes. Despite the fact that memristors are two-terminal devices, we have found that the above can be rigorously achieved by essentially a subtraction procedure: subtracting between data from self-similar Si- memristor device in different resistance states. This procedure is especially convenient in our study because, despite the different Si/Ge film thickness, resistance and composition, all our PPMS, SCM1 and SCM2 measurements used sample cells of the same size and configuration, which are likely to have the same parasitic load resistance due to leads, electrodes, spreading resistance and electrode/film interfaces. The following example illustrates the basic idea. Consider in our study two cells that contain two different films that are self-similar in the following sense: At 10 T, both films have a magnetoresistance that is 1% of its 0 T resistance. (Note that their cell magnetoresistance, which includes the magnetoresistance of both the film and the parasitic resistance, is not self-similar even if film's magnetoresistance is.) If so, then the difference of the two cell-resistances will also show a magnetoresistance of 1% of its 0 T value, because unlike the cell-resistance the difference does not include the parasitic resistance. This example suggests that, instead of analyzing the two-terminal cell resistance individually, we can analyze a set of them, and for each pair within the set, we can find its resistance difference within the pair. Such pair-resistance-differences will carry the same information of the relative resistance change of the film. (There is an analogous case in diffraction of matter: The diffraction intensity is the Fourier transform of the vectors of atom-pairs; it is *not* the Fourier transform of the position vectors of atoms. Nevertheless, the diffraction intensity still informs the structure of matter satisfactorily.) We have formalized this analysis and reported it elsewhere (See ref. 38). Below is a summary of the procedure used for data analysis here.

Consider all the cells have the same parasitic resistance. Let the resistances of two self-similar cells, 1 and 2, increase from $R(1)$ and $R(2)$ at $(T,B)=(0,0)$ to $R'(1)$ to $R'(2)$ at $(T,B)$, respectively. (Here, $R(1)$, $R(2)$ and their difference $R(1)-R(2)$ can be regarded as $R_{0K}$.) Then the relative conductivity change $\Delta\sigma/\sigma$ can be obtained from the relative change of the resistance difference of the cell pair, $-\Delta R/R_0$. Here, $R_0=R(1)-R(2)$ and $\Delta R=(R'(1)-R'(2))-(R(1)-R(2))=(R'(1)-R(1))-(R'(2)-R(2))$ is the change of the resistance difference due to the change of $(T,B)$. The above result is exact. This method can be used to accurately determine $\Delta\sigma/\sigma_0=-\Delta R/R_0$ down to 0.1%.

In our data analysis, for each composition we first selected a set of metallic cells that differ in either Si/Ge thickness or the resistance value. These cells will be named cell 1, 2, etc. Assuming they are self-similar, we followed the resistance difference method to calculate $-\Delta R$. (The resistance difference between two cells used for the following analysis was in the Ohmic regime.) The self-similarity assumption is considered validated if (i) the resultant $-\Delta R$ obeys the same scaling law; for example, $-\Delta R$ of various $(T,0)$ follows the same temperature scaling law, and (ii) $\Delta\sigma/\sigma=-\Delta R/R$ of all the pairs falls on a universal curve/line consistent with the scaling law.

Comparing thus obtained $\Delta\sigma/\sigma_0 = -\Delta R/R_0$ data with a conductivity model then allows determination of the model parameters for each composition.

**Mechanistic model fitting to data-Quantum correction to conductivity (QCC)**: (For more details, see Ref. 21 and Ref. 22.) Coherent electrons in a random system give rise to quantum correction to conductivity (QCC) in three ways. First, their interference as spinless electrons undergoing random walk (i.e., diffusion) gives rise to an enhanced amplitude of the wave function upon back scattering to the origin, yielding increased localization (hence called weak localization or WL), reduced conductivity and increased resistivity. This effect becomes more prominent at lower temperature as the diffusion length $L_T$ increases, so WL resistance increases with decreasing temperature in a dimension-dependent manner. Second, when spin-orbit interaction is strong, the above mechanism must also consider the spin contributions of the two-electron pair, which gives rise to a qualitatively similar correction but with an opposite sign. Therefore, this is the weak anti-localization hence called the WAL mechanism. Third, in a disordered system, diffusing electrons especially those in three dimensions must take a much longer time before they can leave each other after each collision. In essence, two colliding electrons after each encounter are "stuck together" until they find a way to diffuse away by tortuous random walk. There is thus enhanced electron-electron interaction, of which the low energy contribution and, when there is a magnetic field, the Zeeman splitting of the initial and final states, are most important. The resistance correction due to such electron-electron-interaction (EEI) mechanism has the same temperature and dimensionality dependence as the WL mechanism. However, whereas the WL contribution is solely dependent on the diffusion length, and is thus saturated below certain temperature when $L_T$ saturates at the sample boundary, there is no saturation in the EEI contribution because the $k_B T$-normalized Zeeman effect continues to increase with decreasing temperature despite the fact that $L_T$ has saturated. Another distinguishing feature is in magnetoresistance (MR). Whereas WL gives a negative MR because a magnetic field destroys the quantum coherence of an electron traveling back to the origin, hence removing the QCC that increases the resistance, EEI gives a positive MR because the Zeeman effect reduces the final state statistics thus reduces the conductivity. These qualitatively different behaviors allow us to determine the dominant mechanism from the trend of the data.

Specifically, our isotropic magnetoresistance (MR) implies that the disordered system is 3D in nature. Next, given the observations of (i) increased resistance at lower temperature and (ii) positive MR, we can rule out WAL in our memristors. Third, a saturation of QCC was indeed observed, indicating the saturation of $L_T$. However, their MR continues to increase at lower temperature, which can only be explained by EEI. Therefore, it is reasonable to assume EEI is the dominant contribution in our model fitting.

According to the literature[35], the EEI QCC as a function of temperature and magnetic field in 3D is given by

$$\Delta\sigma(T,B)=\sigma(T,B)-\sigma_o=\alpha(4/3-3\tilde{F}_\sigma/2)\sqrt{T}-0.77\alpha\tilde{F}_\sigma\sqrt{T}g_3(\tilde{h}) \qquad (1)$$

Here, $\sigma_o = \sigma(0,0)$ is 0K conductivity which is the uncorrected Drude conductivity at 0K, $B$ is magnetic field causing normalized Zeeman splitting $\tilde{h} = g\mu_B B/k_B T$ (we shall set $g = 2$), and $g_3$ is a known even function of $\tilde{h}$ with asymptotes $\tilde{h}^2$ at $\tilde{h} \ll 1$ and $\tilde{h}^{1/2}$ at $\tilde{h} \gg 1$. Therefore, $\Delta\sigma/\sigma_o$ that equals $-\Delta R/R_0$ depends on only two independent materials parameter: (i) $\alpha/\sigma_o$, where $\alpha = (e^2/\hbar)(1.3/4\pi^2)(k_B/2\hbar D)^{1/2}$, which is set by electron diffusivity $D$, and (ii) Fermi-liquid constant $\tilde{F}_\sigma$, which specifies the strength and sign of EEI. Note that the second term above is the MR, while the first term is the zero-field QCC. Note also that the $\sqrt{T}$ dependence comes from the temperature dependence of $L_T$, and below the saturation temperature it should be set as $\sqrt{T_{saturation}}$. Therefore, having both MR and the zero-field QCC allows the unique determination of the two independent material parameters, and the goodness of the fit serves to verify the validity of the mechanism. This is shown in **Fig. 1d**, **Fig. 4a** and **Supplementary Figure 9** for the P-Si:0.45N memristors in the LRS (**Supplementary Figure 9** is the same as **Fig. 1d**, which shows a good fit), including ones with a film thickness from 8 nm to 16 nm (**Fig. 4a**), and for the 11 nm memristors, set in various intermediate states from 28 Ω to 393 Ω and at different temperatures (**Supplementary Figure 9**). Therefore, these results validate not only the EEI mechanism but also the self-similarity of their conductivity, in that they all share a universal $\Delta\sigma/\sigma_o$, thus they must have the same $\alpha/\sigma_o$ and $\tilde{F}_\sigma$. Altogether, this analysis was applied to 45 sets of data from the P-Si:0.45N memristors series, all of them fit by the same set of two independent materials parameters above. So, all such memristors must have similar conducting networks that share the same model characteristics, and they can at most differ in network density. Other than the P-Si:0.45N memristors, we have also applied the same analysis to LRS memristors of different compositions with same success in establishing the validity of Eq. (1) and the universality. **Supplementary Table S1** summarizes the model parameters for some memristors from this analysis, and further discussion of these results can be found elsewhere[22].

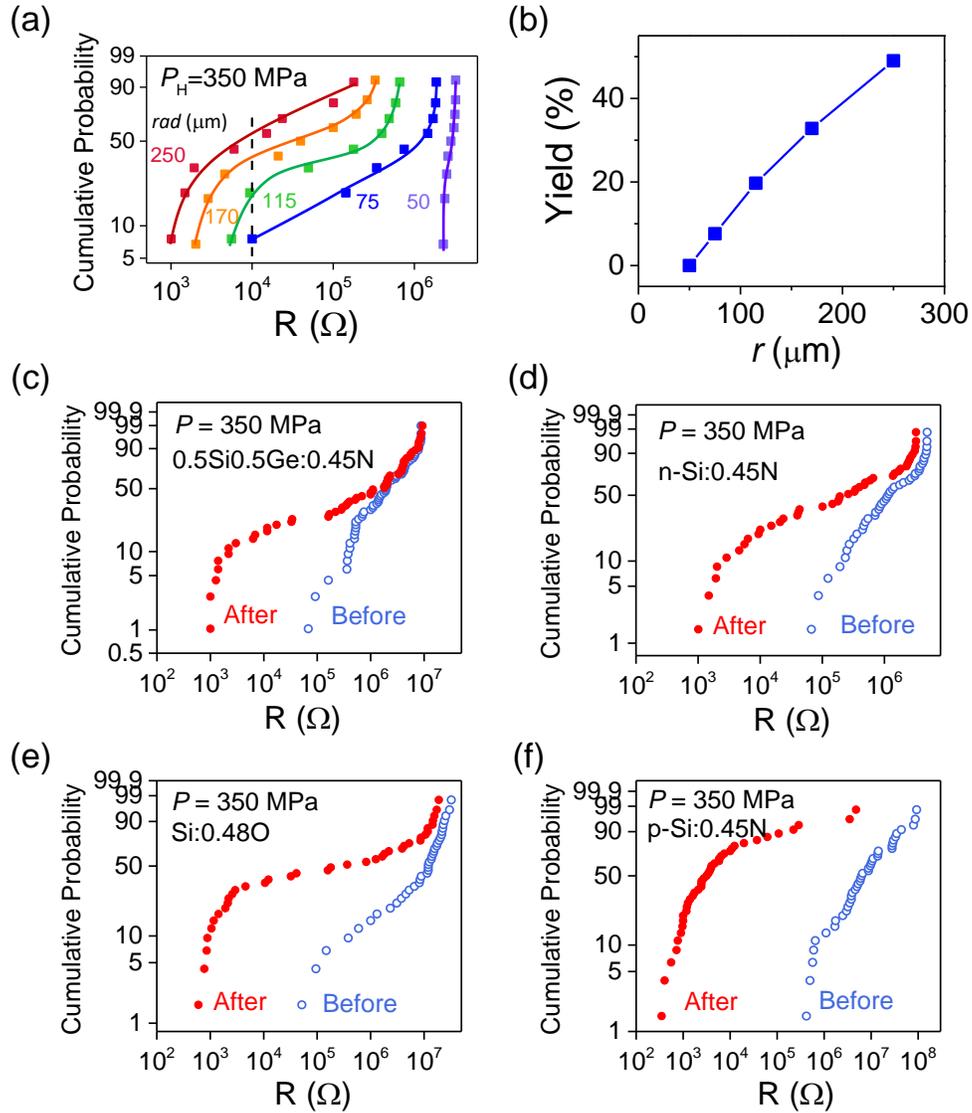

**Figure S1** (a)-(b): Size dependence of pressure switching in Mo/SiO$_{0.25}$/Pt memristor (thickness: 10 nm). (a) Weibull plots of resistance distribution of 10 cells with various device radii *rad* after applying $P_H$=350 MPa. All cells preset to HRS before applying pressure. (b) Switching yield, defined by threshold resistance value of $10^4$ Ω (dashed line in (a)), increases with area. Solid lines are guide to the eyes. (c)-(f): HRS-to-LRS transition induced by $P_H$= 350 MPa in various O/N-doped Si/Ge memristors: (c) Si0.5Ge$_{0.5}$N$_{0.45}$, (d) P-Si N$_{0.45}$, (e) SiO$_{0.48}$, (f) B-Si N$_{0.45}$. All with 10 nm thickness and 250 μm cell radius.

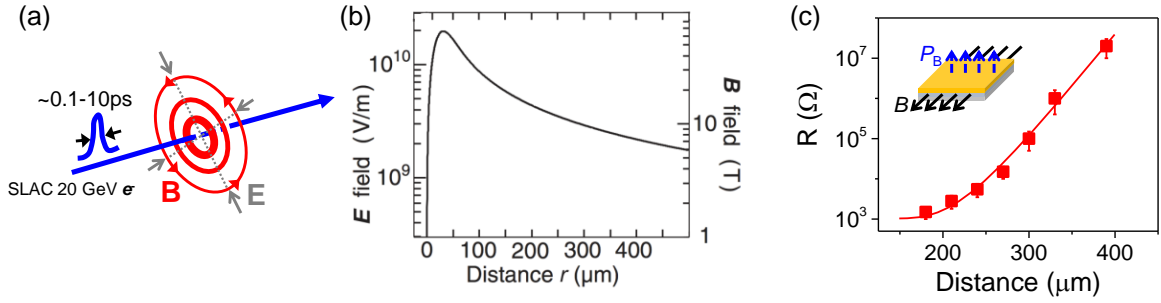

**Figure S2** (a) Electron bunch made of 20 GeV electrons bunched into a time packet of 0.1-10 ps (a) generates concentric magnetic field *B* shown in red. (b) (From **Ref. S1**.) Pulse magnetic field lasting $10^{-13}$ s generated by electron bunch peaks at the perimeter (40 μm) of the bunch. Its far field follows $B \sim 1/distance$ and magnetic pressure follows $P_B \sim 1/(distance)^2$ where *distance* is radial from the travel path of the bunch. (c) P-Si $N_{0.45}$ memristor cells, all pre-switched to HRS $\sim 10^8$ Ω, were switched to LRS by $P_B$ (inset) with decreasing resistance at decreasing distance to the center (resistance read at 0.2 V). Lower bound estimate for $P_B$: *distance*=40 μm (electron-bunch's perimeter), ~1,950 MPa; *distance*=180 μm, ~96 MPa; *distance*=390 μm (outer end of the LRS region), ~21 MPa. See more details in Ref. S2-S3.

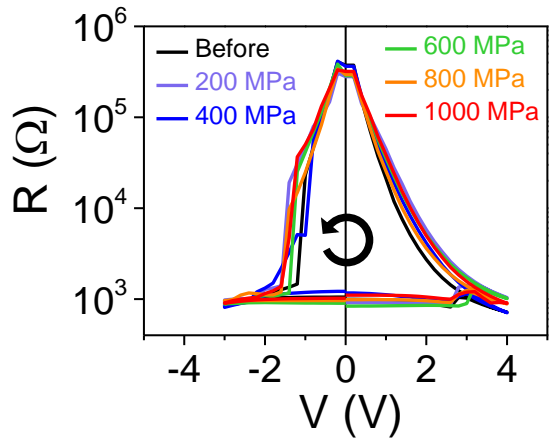

**Figure S3** Similar *R-V* switching curves before and after pressure switching (from HRS to LRS) by different hydraulic pressure as marked. Mo/SiO$_{0.25}$/Pt devices with 10 nm thickness and 250 μm cell radius.

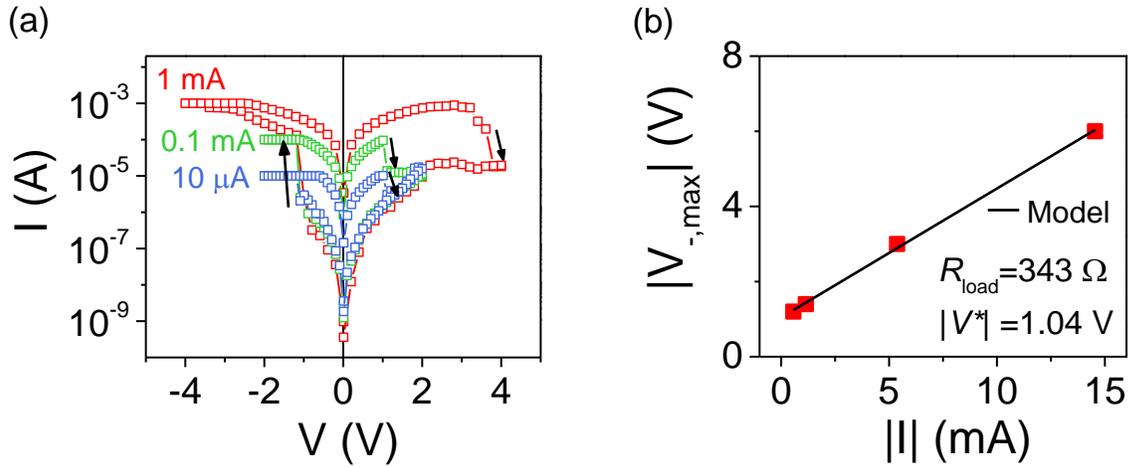

**Figure S4** Multiple LRS resistance in the same Si-memristor obtained by varying SET current compliance $I_{cc}$. (a) $I$-$V$ curves of P-Si N$_{0.45}$ devices with various $I_{cc}$. The higher the $I_{cc}$, the lower the $R_{LRS}$, the higher the positive RESET voltage. Arrows indicate switching direction. (b) Linear relation between $|V_{-,max}|$ and $|I|$ as predicted by $|V_{-,max}|$ =maximum negative-polarity voltage = $V^* + IR_{load}$ per model of **Fig. 2a** in the main text. Intercept and slope give $V^*$=1.04 V and $R_{load}$=343 Ω, respectively.

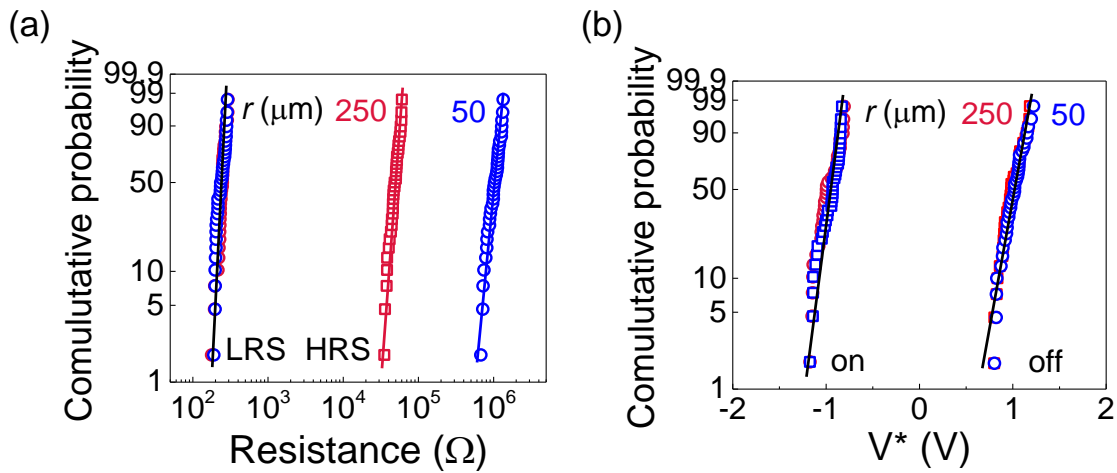

**Figure S5** Device uniformity of Mo/SiO$_{0.25}$/Pt memristors (thickness: 11 nm) in two sizes (device radius =250 μm and 50 μm) represented in Weibull plots of (a) HRS and LRS resistance (load resistance $R_{load}$ excluded), and (b) critical switching voltage $V^*$. Each distribution contains 35 devices tested under the same voltage-sweeping condition. The extracted standard-deviation-to-mean $\Delta/\mu$ and Weibull slope $k$ of HRS/LRS resistance and switching voltage $V^*$ are plotted in **Fig. 2f** in comparison with literature data.

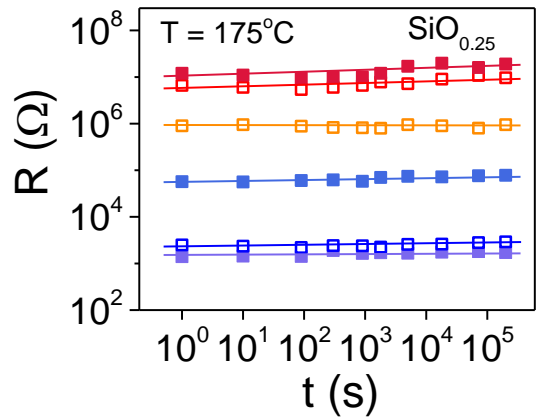

**Figure S6** Data retention of Mo/SiO$_{0.25}$/Pt memristors over $10^5$ s at 175 °C for multiple resistance states obtained by DC voltage switching (filled symbols) and 100 ns pulse switching (hollow symbols).

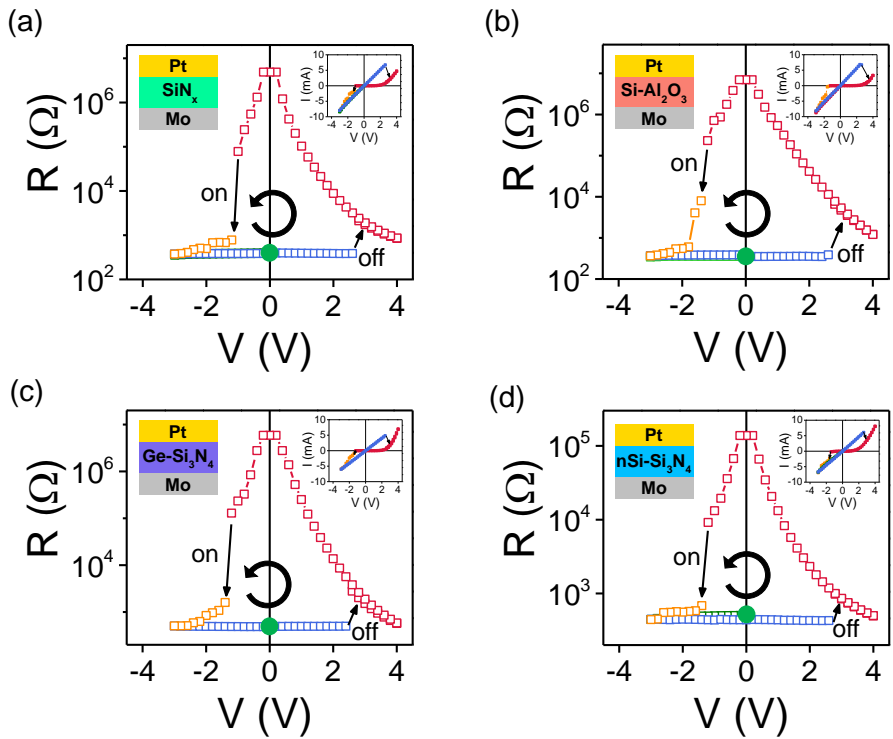

**Figure S7** Representative switching curves with almost identical *R-V* hysteresis for memristors made of (a) Si+N$_2$, (b) Si+Al$_2$O$_3$, (c) Ge+Si$_3$N$_4$, and (d) P-Si+Si$_3$N$_4$. Device schematics and corresponding *I-V* curves are shown as insets. Counterclockwise arrows indicate switching directions. All devices have ~10 nm thickness and 100 μm radius. Each test used a virgin device never tested before, with an initial resistance indicated by the oversized green circle.

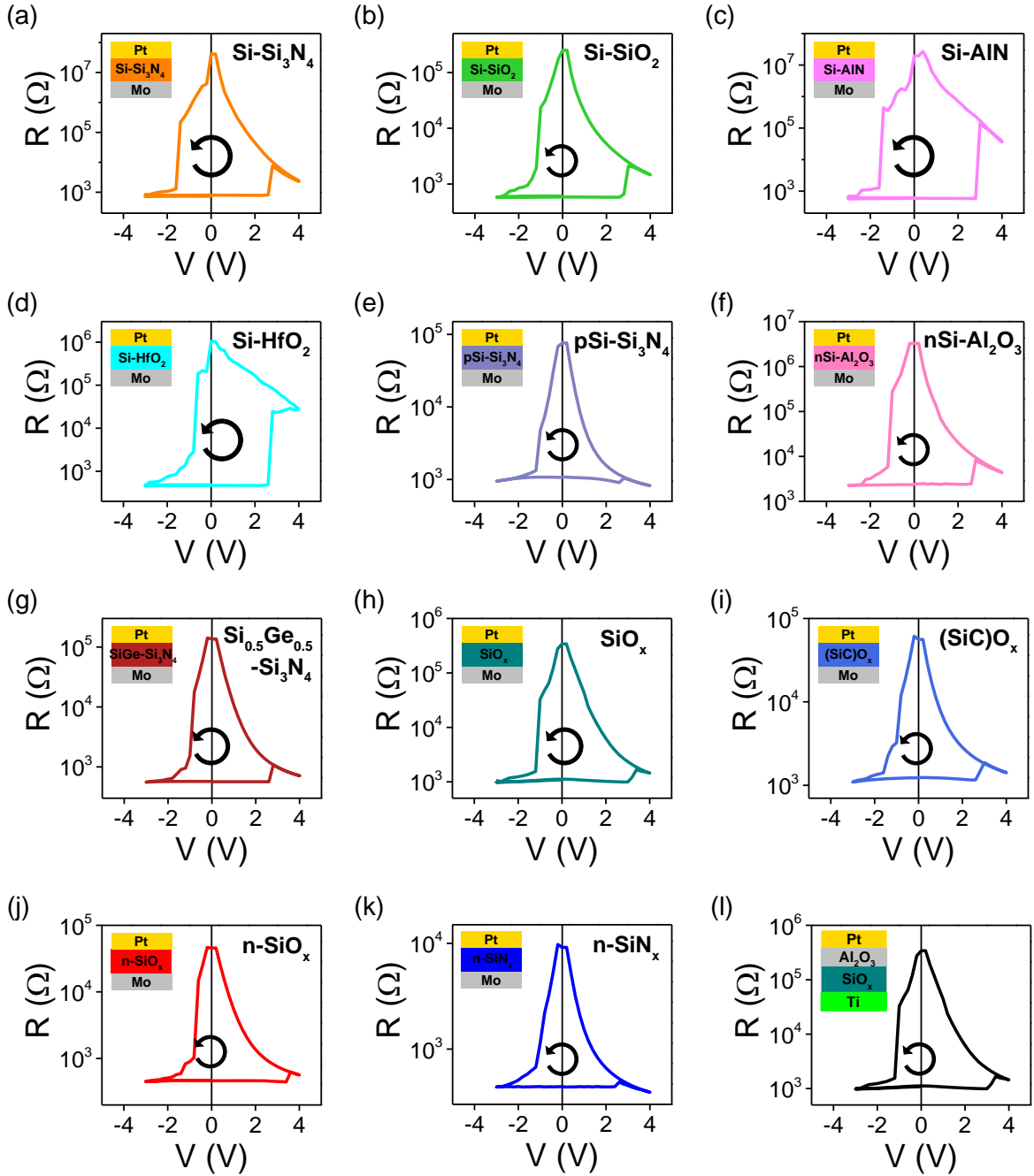

**Figure S8** Memristors of various O/N-doped Si/Ge compositions have similar *R-V* hysteresis and switching voltage. (a) Si+$Si_3N_4$, (b) Si+$SiO_2$, (c) Si+AlN, (d) Si+$HfO_2$, (e) B-Si+$Si_3N_4$, (f) B-Si+$Al_2O_3$, (g) $Si_{0.5}Ge_{0.5}$ +$Si_3N_4$, (h) Si+$O_2$, (i) SiC+$O_2$, (j) P-Si+$O_2$, (k) B-Si+$N_2$, and (l) pure Si with additional layer of $Al_2O_3$ that oxidized Si into $SiO_x$. Insets illustrate schematic device structure with Pt as top electrode and Mo as bottom electrode in (a)-(k), and Mo replaced by Ti in (l). Counterclockwise arrows indicate same switching direction. All devices have ~ 10 nm thickness and 100 μm cell radius.

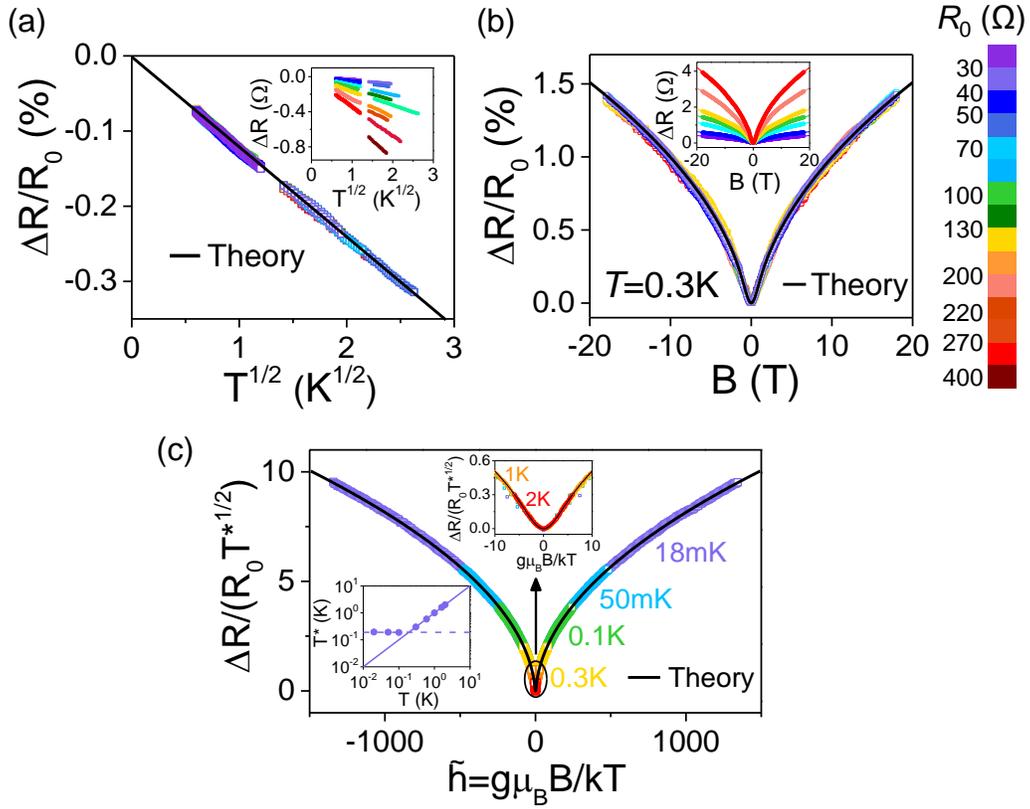

**Figure S9** (a)-(b): Universal QCC shared by multiple LRS of 11 nm P-Si:0.45N memristor as demonstrated by overlapped relative resistance change $\Delta R$ against (a) $T^{1/2}$ below $T_{min}$, and (b) magnetic field at 0.3K. Insets: plots of $\Delta R$ of individual metallic states before normalization. Solid lines: predictions by Eq. (1) in **Methods**. Various resistance states with $R_0$ indicated by color spectrum on right. (c) Normalized magnetoresistance vs. $\tilde{h} = g\mu_B B/k_B T$ of 11 nm P-Si:0.45N memristor at different temperatures (as labeled). Relative MR divided by $T^{*1/2}$ (see lower inset) has overlapping dependence, varying with $B^{1/2}$ at high field and with $B^2$ at low field (upper inset), consistent with prediction by Eq. (1) in **Methods** shown as solid curves.

**Table S1** Model parameters according to the analysis using Eq. (1) described in **Methods**. Various devices made from same composition have the same parameters.

| Composition | $\tilde{F}_\sigma$ | $\sigma_o D^{1/2}$ $(\Omega \cdot s^{1/2})^{-1}$ |
|---|---|---|
| Si | 0.34 | 289 |
| n-Si | 0.48 | 352 |
| n-Si:0.45N | 0.75 | 346 |
| Si:0.48O | 0.65 | 351 |